\RequirePackage{amsmath}
\documentclass[runningheads]{llncs}
\usepackage[T1]{fontenc}
\usepackage{graphicx}
\usepackage{hyperref}
\usepackage{color}

\usepackage[caption=false]{subfig}
\usepackage{tikz}
\usepackage{amssymb}
\usepackage{clrscode3e}
\usepackage{setspace}
\begin{document}
\title{Nonogram: Complexity of Inference and Phase Transition Behavior}
\author{Aaron Foote\inst{1} \and
Danny Krizanc\inst{1}}
\authorrunning{A. Foote et al.}
\institute{Wesleyan University, Middletown CT 06459, USA \\ \email{\{afoote,dkrizanc\}@wesleyan.edu}}
\maketitle              % typeset the header of the contribution
\begin{abstract}
Nonogram is a popular combinatorial puzzle similar in nature to Sudoku or Minesweeper in which a puzzle solver must determine if there exists a setting of the puzzle parameters that satisfy a given set of constraints. It has long been known that the problem of deciding if a solution exists is NP-complete. Despite this fact, humans still seem to enjoy playing it. This work aims to reconcile these seemingly contradictory facts by (1) analyzing the complexity of the inference problem for Nonogram (if there exists an unfilled cell whose value is forced) and (2) experimentally establishing the existence of a phase transition behavior for this inference problem. Our results show that the difficulty of the inference problem is largely determined by the density of filled cells (positive parameters) in a given puzzle. Along the way we implement an efficient encoding of a Nonogram board as a Boolean formula in Conjunctive Normal Form (CNF) through the use of regular expressions in order to make our experiments feasible.

\keywords{Nonogram \and Combinatorial Puzzles  \and Phase Transition \and Complexity of Games}
\end{abstract}
\section{Introduction}

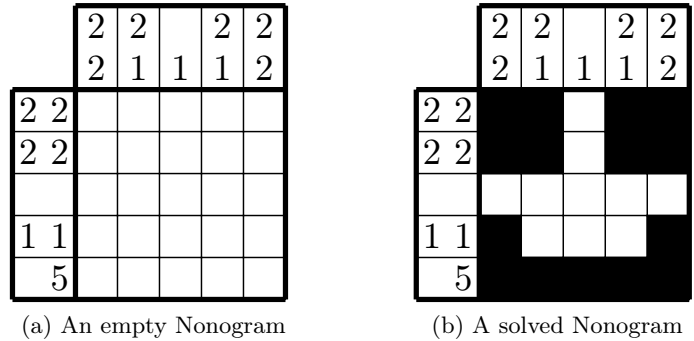
\begin{figure}
    \centering
    \subfloat[An empty Nonogram]{\resizebox{0.32\textwidth}{!}{\begin{tikzpicture}           
    \draw[step=0.4cm] (0,0) grid (2,2);
                
    \draw[line width=0.5mm] (-0.6,0) -- (2,0);
    \draw[line width=0.5mm] (-0.6,2) -- (2,2);
    \draw[line width=0.5mm] (-0.6,0) -- (-0.6,2);
    
    \draw (0,0.4) -- (-0.6,0.4);
    \draw (0,0.8) -- (-0.6,0.8);
    \draw (0,1.2) -- (-0.6,1.2);
    \draw (0,1.6) -- (-0.6,1.6);
    \draw (0,2.0) -- (-0.6,2.0);

    \node at (-0.15,0.2) {5};
    
    \node at (-0.15,0.6) {1};
    \node at (-0.45,0.6) {1};

    \node at (-0.15,1.4) {2};
    \node at (-0.45,1.4) {2};
    
    \node at (-0.15,1.8) {2};
    \node at (-0.45,1.8) {2};
    
    \draw[line width=0.5mm] (0,0) -- (0,2.8);
    \draw[line width=0.5mm] (2,0) -- (2,2.8);
    \draw[line width=0.5mm] (0,2.8) -- (2,2.8);
    
    \draw (0.4,2) -- (0.4,2.8);
    \draw (0.8,2) -- (0.8,2.8);
    \draw (1.2,2) -- (1.2,2.8);
    \draw (1.6,2) -- (1.6,2.8);
    \draw (2.0,2) -- (2.0,2.8);
    
    \node at (0.2,2.2) {2};
    \node at (0.2,2.6) {2};

    \node at (0.6,2.2) {1};
    \node at (0.6,2.6) {2};

    \node at (1.0,2.2) {1};
    
    \node at (1.4,2.2) {1};
    \node at (1.4,2.6) {2};
 
    \node at (1.8,2.2) {2};
    \node at (1.8,2.6) {2};
\end{tikzpicture}}}\hfil
    \subfloat[A solved Nonogram]{\resizebox{0.32\textwidth}{!}{\begin{tikzpicture}           
    \draw[step=0.4cm] (0,0) grid (2,2);
                
    \draw[line width=0.5mm] (-0.6,0) -- (2,0);
    \draw[line width=0.5mm] (-0.6,2) -- (2,2);
    \draw[line width=0.5mm] (-0.6,0) -- (-0.6,2);
    
    \draw (0,0.4) -- (-0.6,0.4);
    \draw (0,0.8) -- (-0.6,0.8);
    \draw (0,1.2) -- (-0.6,1.2);
    \draw (0,1.6) -- (-0.6,1.6);
    \draw (0,2.0) -- (-0.6,2.0);

    \node at (-0.15,0.2) {5};
    
    \node at (-0.15,0.6) {1};
    \node at (-0.45,0.6) {1};

    \node at (-0.15,1.4) {2};
    \node at (-0.45,1.4) {2};
    
    \node at (-0.15,1.8) {2};
    \node at (-0.45,1.8) {2};
    
    \draw[line width=0.5mm] (0,0) -- (0,2.8);
    \draw[line width=0.5mm] (2,0) -- (2,2.8);
    \draw[line width=0.5mm] (0,2.8) -- (2,2.8);
    
    \draw (0.4,2) -- (0.4,2.8);
    \draw (0.8,2) -- (0.8,2.8);
    \draw (1.2,2) -- (1.2,2.8);
    \draw (1.6,2) -- (1.6,2.8);
    \draw (2.0,2) -- (2.0,2.8);

    \filldraw[draw=black,color=black] (0,0) rectangle (2.0,0.4);
    \filldraw[draw=black,color=black] (0,0.4) rectangle (0.4,0.8);
    \filldraw[draw=black,color=black] (1.6,0.4) rectangle (2.0,0.8);
            
    \filldraw[draw=black,color=black] (0,1.2) rectangle (0.8,2.0);
    \filldraw[draw=black,color=black] (1.2,1.2) rectangle (2.0,2.0);
    
    \node at (0.2,2.2) {2};
    \node at (0.2,2.6) {2};

    \node at (0.6,2.2) {1};
    \node at (0.6,2.6) {2};

    \node at (1.0,2.2) {1};
    
    \node at (1.4,2.2) {1};
    \node at (1.4,2.6) {2};
 
    \node at (1.8,2.2) {2};
    \node at (1.8,2.6) {2};
\end{tikzpicture}}}\hfil
    \caption{An example Nonogram of a smiley face.}
    \label{fig:Smiley}
\end{figure}

Nonogram is a combinatorial puzzle game played on an initially empty grid of cells, with some filled in for the solution. Each row and column is associated with a sequence of integers specifying lengths of runs of filled cells in that row/column, termed a \emph{description}. Players form a solution by assigning a subset of the cells to be filled such that every row and column description is satisfied. A unique aspect of Nonogram is that the solutions typically form a recognizable image (Figure \ref{fig:Smiley}).

Due to its popularity, Nonogram is a well-studied puzzle. Prior work shows the problem of consistency (the existence of a solution) NP-complete through parsimonious reduction \cite{parsimonious} as well as with tools more specialized for combinatorial puzzles \cite{Hearn,NonogramNPComplete1,NonogramNPComplete2}. Various algorithmic approaches exist for solving Nonogram, including regular expression matching \cite{Bosboom}, evolutionary algorithms \cite{ev}, and integer programming \cite{integerProgramming,constraintProgramming}, along with a slew of heuristic approaches (see \href{https://webpbn.com/survey/}{https://webpbn.com/survey/} for an extensive review of solvers). 

\begin{figure}[h]
    \centering
    \captionsetup{justification=centering}
    \subfloat[Inference possible with multiple solutions]{\resizebox{0.32\textwidth}{!}{\begin{tikzpicture}           
    \draw[step=0.4cm] (0,0) grid (2.4,2.4);
                
    \draw[line width=0.5mm] (-0.45,0) -- (2.4,0);
    \draw[line width=0.5mm] (-0.45,2.4) -- (2.4,2.4);
    \draw[line width=0.5mm] (-0.45,0) -- (-0.45,2.4);
    
    \draw (0,0.4) -- (-0.45,0.4);
    \draw (0,0.8) -- (-0.45,0.8);
    \draw (0,1.2) -- (-0.45,1.2);
    \draw (0,1.6) -- (-0.45,1.6);
    \draw (0,2.0) -- (-0.45,2.0);

    \node at (-0.15,0.2) {1};
    \node at (-0.15,1.0) {1};
    \node at (-0.15,1.8) {2};
    \node at (-0.15,2.2) {2};
                
    \draw[line width=0.5mm] (0,0) -- (0,3);
    \draw[line width=0.5mm] (2.4,0) -- (2.4,3);
    \draw[line width=0.5mm] (0,3) -- (2.4,3);
    
    \draw (0.4,2.4) -- (0.4,3);
    \draw (0.8,2.4) -- (0.8,3);
    \draw (1.2,2.4) -- (1.2,3);
    \draw (1.6,2.4) -- (1.6,3);
    \draw (2.0,2.4) -- (2.0,3);
    
    \node at (0.2,2.6) {2};
    \node at (0.6,2.6) {2};
                 
    \node at (1.4,2.6) {1};
 
    \node at (2.2,2.6) {1};
\end{tikzpicture}}}\hfil
    \subfloat[Inference possible with a unique solution]{\resizebox{0.32\textwidth}{!}{\begin{tikzpicture}           
    \draw[step=0.4cm] (0,0) grid (2.4,2.4);
                
    \draw[line width=0.5mm] (-0.45,0) -- (2.4,0);
    \draw[line width=0.5mm] (-0.45,2.4) -- (2.4,2.4);
    \draw[line width=0.5mm] (-0.45,0) -- (-0.45,2.4);
    
    \draw (0,0.4) -- (-0.45,0.4);
    \draw (0,0.8) -- (-0.45,0.8);
    \draw (0,1.2) -- (-0.45,1.2);
    \draw (0,1.6) -- (-0.45,1.6);
    \draw (0,2.0) -- (-0.45,2.0);

    \node at (-0.15,0.2) {1};
    \node at (-0.15,1.8) {2};
    \node at (-0.15,2.2) {2};
                
    \draw[line width=0.5mm] (0,0) -- (0,3);
    \draw[line width=0.5mm] (2.4,0) -- (2.4,3);
    \draw[line width=0.5mm] (0,3) -- (2.4,3);
    
    \draw (0.4,2.4) -- (0.4,3);
    \draw (0.8,2.4) -- (0.8,3);
    \draw (1.2,2.4) -- (1.2,3);
    \draw (1.6,2.4) -- (1.6,3);
    \draw (2.0,2.4) -- (2.0,3);
    
    \node at (0.2,2.6) {2};
    \node at (0.6,2.6) {2};
 
    \node at (2.2,2.6) {1};
\end{tikzpicture}}}
    \caption{Two puzzles in which inference is possible (the 2x2 square). In the first puzzle the pair of 1 clues creates multiple solutions, while the second puzzle only has one solution.}
    \label{fig:InferenceIsNotUniqueness}
\end{figure}
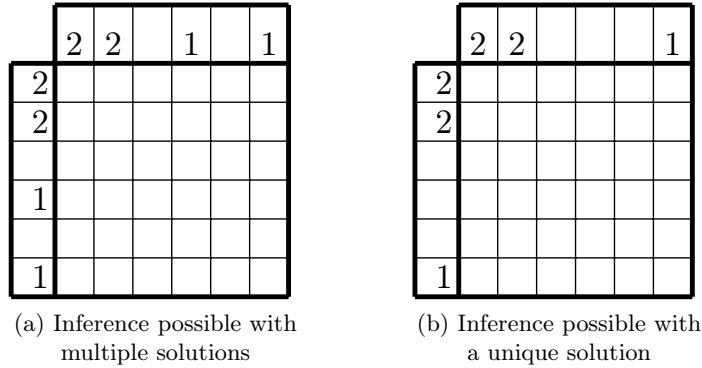

As noted by Kaye in the context of Minesweeper \cite{MinesweeperNPC}, obtaining an NP-complete result for consistency does not guarantee that there is not a polynomial-time approach to playing the game. Later work proposes the problem of inference -- deciding the existence of a cell that can have its value deduced by logical inference -- as a more natural expression of solving a combinatorial puzzle. After all, when players are handed a puzzle to play, they typically assume that there is a solution (consistency) and only one of them (uniqueness). The player is really interested in how difficult it is to make progress, and if any progress can be made at all without guessing, which is the inference problem. This problem has been shown co-NP-complete for Minesweeper \cite{InferenceCNPC}. Importantly, the inference problem is \emph{not} the complement of the uniqueness problem which is known to be NP-complete \cite{NonogramNPComplete1}. An inference instance can be positive for puzzles with and without unique solutions (Figure~\ref{fig:InferenceIsNotUniqueness}). It is not hard to show that a puzzle has a unique solution if and only if every cell is inferable \cite{MYthesis}. Furthermore, a satisfiable puzzle without any cells assigned may not have an inference possible. 

The study of phase transition behavior in combinatorial problems is useful for understanding the average-case complexity of problems that are infeasible for humans to solve in the worst
case. With this, a puzzle maker can tune their algorithm to generate puzzles that require more or less steps to solve. Typically, a parameter of the combinatorial problem can be identified as a threshold, on each side of which a characteristic of the problem is dramatically different. The ratio of clauses to variables is such a parameter for Boolean satisfiability \cite{SATpt}, and edge density a similar parameter for k-colorability of graphs \cite{GraphColoringPT}. For the puzzles of Sudoku and Minesweeper, phase transition behavior for the consistency problem has been shown \cite{MinesweeperPT,SudokuPT}.

In section \ref{sec:Inference co-NP-complete}, the problem of inference for Nonogram is defined and shown to be co-NP-complete. In section \ref{sec:Phase Transition}, phase transition behavior for the inference problem is demonstrated empirically. In the process of demonstrating this phase transition, a reduction from Nonogram consistency to CNF formulae is developed, leveraging the expressibility of Nonogram descriptions as regular expressions. The formulas can be solved iteratively under different assumptions to solve the inference problem.

\section{Inference is co-NP-complete}
\label{sec:Inference co-NP-complete}

The INFERENCE problem for Nonogram is defined as follows:

\paragraph{INSTANCE:} A consistent Nonogram board $N = (R,C,m,n)$ and mapping of cells to values $f$, where $R$ is the row descriptions and $C$ the column descriptions -- each a sequence of integer sequences -- and $m$ and $n$ the dimensions of the board.
\paragraph{QUESTION:} Does there exist a cell $c$ such that for all solutions $f$ to $N$, that $f(c) = v$, i.e., is $c$ mapped to the same value in every solution?\\
%\[\text{INFERENCE} = \{\langle N \rangle \vert N \text{ is a Nonogram board with a cell that can have its value inferred}\}.\]

Here we show that the INFERENCE problem for Nonogram is co-NP-complete. To this end, it is first shown that negative instances for INFERENCE can be verified in polynomial time. Next, the INFERENCE problem is proved co-NP-hard by a reduction from Boolean unsatisfiability.

\begin{theorem}\label{th:INFERENCE in co-NP}
    INFERENCE is co-NP complete.
\end{theorem}
\begin{proof}
    We first show INFERENCE is a member of co-NP. For an $m \times n$ Nonogram board $N$, a certificate for a negative instance to the INFERENCE problem is the following: for each cell $c_{ij}$ two consistent fillings of $N$ are presented such that one has $c_{ij}$ filled and the other $c_{ij}$ empty. Thus, no cell can have its value inferred. A filled board can be checked for consistency by simply verifying that each description is satisfied. There are $2mn$ boards to check and $m + n$ descriptions for each board, thus the certificate can be verified in time polynomial in the size of the Nonogram board.

    Now, we show it co-NP-hard. We prove INFERENCE co-NP-hard through a reduction from Boolean unsatisfiability. First, modify the input formula $\phi$ to be $\psi = \phi \land x^*$, where 
    $x^*$ is a variable that does not occur in $\phi$. Note that if $\phi$ is satisfiable, so is $\psi$ (take $x^*$ to be true). Further, if $\phi$ is unsatisfiable, then $\psi$ is 
    unsatisfiable. We can then construct Boolean circut $C_{\psi}$ that is unsatisfiable if and only if $\psi$ is unsatisfiable. Below are Nonogram gadgets for circuit components of AND, 
    OR, NOT, input/output, wires, joints, and splitters, and crossovers (Figure \ref{fig:Gadgets}). More details for the gadgets, including their solutions, can be found in \cite{MYthesis}. All gadgets have the following properties:
    \begin{enumerate}
        \item There is at least one pair of input cells corresponding to an input variable (with the exception of input/output terminals).
        \item The input cells align with the output cells of the preceding gadget, and output cells align with the input cells of the gadget that follows.
        \item The positive input cell ($v$) is filled if and only if the negative output cell ($\Bar{v}'$) of the preceding gadget is filled.
        \item It is impossible to derive the contents of any gadget without information from other gadgets (with the exception of input terminals).
        \item All gadgets are consistent, meaning they have a solution.
        \item Each gadget has eleven rows and eleven columns.
    \end{enumerate}

    \begin{figure}[ht!]
        \centering
            \subfloat[NOT gadget\label{fig:NOT Gadget}]{\resizebox{0.32\textwidth}{!}{\begin{tikzpicture}
    \node at (0.2,2.2) {\textcolor{black}{$\Bar{v}$}};
    \node at (1,2.2) {\textcolor{black}{$v$}};
    \node at (4.2,2.2) {\textcolor{black}{$\Bar{v}'$}};
    \node at (3.4,2.2) {\textcolor{black}{$v'$}};
                
    \draw[step=0.4cm] (0,0) grid (4.4,4.4);
                
    \draw[line width=0.5mm] (-0.9,0) -- (4.4,0);
    \draw[line width=0.5mm] (-0.9,4.4) -- (4.4,4.4);
    \draw[line width=0.5mm] (-0.9,0) -- (-0.9,4.4);
    
    \draw (0,0.4) -- (-0.9,0.4);
    \draw (0,0.8) -- (-0.9,0.8);
    \draw (0,1.2) -- (-0.9,1.2);
    \draw (0,1.6) -- (-0.9,1.6);
    \draw (0,2.0) -- (-0.9,2.0);
    \draw (0,2.4) -- (-0.9,2.4);
    \draw (0,2.8) -- (-0.9,2.8);
    \draw (0,3.2) -- (-0.9,3.2);
    \draw (0,3.6) -- (-0.9,3.6);
    \draw (0,4.0) -- (-0.9,4.0);
    \node at (-0.15,0.2) {1};
    \node at (-0.45,0.2) {1};
    \node at (-0.30,0.6) {11};
    \node at (-0.15,1.0) {1};
    \node at (-0.45,1.0) {1};
    \node at (-0.15,1.4) {2};
    \node at (-0.45,1.4) {4};
    \node at (-0.75,1.4) {2};
    \node at (-0.15,1.8) {1};
    \node at (-0.45,1.8) {2};
    \node at (-0.15,2.2) {2};
    \node at (-0.45,2.2) {2};
    \node at (-0.15,2.6) {2};
    \node at (-0.45,2.6) {2};
    \node at (-0.15,3.0) {2};
    \node at (-0.45,3.0) {4};
    \node at (-0.75,3.0) {2};
    \node at (-0.15,3.4) {1};
    \node at (-0.45,3.4) {1};
    \node at (-0.3,3.8) {11};
    \node at (-0.15,4.2) {1};
    \node at (-0.45,4.2) {1};
                
    \draw[line width=0.5mm] (0,0) -- (0,6);
    \draw[line width=0.5mm] (4.4,0) -- (4.4,6);
    \draw[line width=0.5mm] (0,6) -- (4.4,6);
    \draw (0.4,4.4) -- (0.4,6);
    \draw (0.8,4.4) -- (0.8,6);
    \draw (1.2,4.4) -- (1.2,6);
    \draw (1.6,4.4) -- (1.6,6);
    \draw (2.0,4.4) -- (2.0,6);
    \draw (2.4,4.4) -- (2.4,6);
    \draw (2.8,4.4) -- (2.8,6);
    \draw (3.2,4.4) -- (3.2,6);
    \draw (3.6,4.4) -- (3.6,6);
    \draw (4,4.4) -- (4,6);
    \node at (0.2,4.6) {1};
    \node at (0.2,5) {1};
    \node at (0.2,5.4) {2};
    \node at (0.2,5.8) {1};
    \node at (0.6,4.6) {11};
                 
    \node at (1.0,4.6) {1};
    \node at (1.0,5) {1};
    \node at (1.0,5.4) {1};
    \node at (1.0,5.8) {1};
 
    \node at (1.4,4.6) {1};
    \node at (1.4,5) {1};
    \node at (1.4,5.4) {1};
 
    \node at (1.8,4.6) {1};
    \node at (1.8,5) {1};
    \node at (1.8,5.4) {1};
    \node at (1.8,5.8) {1};
 
    \node at (2.2,4.6) {1};
    \node at (2.2,5) {1};
    \node at (2.2,5.4) {1};
    \node at (2.2,5.8) {1};
 
    \node at (2.6,4.6) {1};
    \node at (2.6,5) {1};
    \node at (2.6,5.4) {1};
    \node at (2.6,5.8) {1};
 
    \node at (3.0,4.6) {1};
    \node at (3.0,5) {1};
    \node at (3.0,5.4) {1};
 
    \node at (3.4,4.6) {1};
    \node at (3.4,5) {1};
    \node at (3.4,5.4) {1};
    \node at (3.4,5.8) {1};
 
    \node at (3.8,4.6) {11};
                 
    \node at (4.2,4.6) {1};
    \node at (4.2,5) {1};
    \node at (4.2,5.4) {1};
    \node at (4.2,5.8) {1};
\end{tikzpicture}}}\hfil
            \subfloat[AND gadget\label{fig:AND Gadget}]{\resizebox{0.32\textwidth}{!}{\begin{tikzpicture}
    \node at (0.2,2.2) {\textcolor{black}{$\Bar{v}_1$}};
    \node at (1.4,2.2) {\textcolor{black}{$v_1$}};
    \node at (4.2,2.2) {\textcolor{black}{$\Bar{v}'$}};
    \node at (3,2.2) {\textcolor{black}{$v'$}};

    \node at (2.2,0.2) {\textcolor{black}{$\Bar{v}_2$}};
    \node at (2.2,1.4) {\textcolor{black}{$v_2$}};

    \draw[step=0.4cm] (0,0) grid (4.4,4.4);

    \draw[line width=0.5mm] (-1.2,0) -- (4.4,0);
    \draw[line width=0.5mm] (-1.2,4.4) -- (4.4,4.4);
    \draw[line width=0.5mm] (-1.2,0) -- (-1.2,4.4);
    \draw (0,0.4) -- (-1.2,0.4);
    \draw (0,0.8) -- (-1.2,0.8);
    \draw (0,1.2) -- (-1.2,1.2);
    \draw (0,1.6) -- (-1.2,1.6);
    \draw (0,2.0) -- (-1.2,2.0);
    \draw (0,2.4) -- (-1.2,2.4);
    \draw (0,2.8) -- (-1.2,2.8);
    \draw (0,3.2) -- (-1.2,3.2);
    \draw (0,3.6) -- (-1.2,3.6);
    \draw (0,4.0) -- (-1.2,4.0);
    \node at (-0.15,0.2) {1};
    \node at (-0.45,0.2) {1};
    \node at (-0.75,0.2) {1};
    \node at (-0.30,0.6) {11};
    \node at (-0.15,1.0) {1};
    \node at (-0.45,1.0) {2};
    \node at (-0.75,1.0) {1};
    \node at (-0.15,1.4) {1};
    \node at (-0.45,1.4) {1};
    \node at (-0.75,1.4) {1};
    \node at (-1.05,1.4) {1};
    \node at (-0.15,1.8) {1};
    \node at (-0.45,1.8) {1};
    \node at (-0.75,1.8) {1};
    \node at (-1.05,1.8) {1};
    \node at (-0.15,2.2) {3};
    \node at (-0.45,2.2) {2};
    \node at (-0.75,2.2) {3};
    \node at (-0.15,2.6) {3};
    \node at (-0.45,2.6) {1};
    \node at (-0.75,2.6) {3};
    \node at (-0.15,3.0) {1};
    \node at (-0.45,3.0) {1};
    \node at (-0.75,3.0) {1};
    \node at (-1.05,3.0) {1};
    \node at (-0.15,3.4) {1};
    \node at (-0.45,3.4) {1};
    \node at (-0.3,3.8) {11};
    \node at (-0.15,4.2) {1};
    \node at (-0.45,4.2) {1};
    
    \draw[line width=0.5mm] (0,0) -- (0,6);
    \draw[line width=0.5mm] (4.4,0) -- (4.4,6);
    \draw[line width=0.5mm] (0,6) -- (4.4,6);
    \draw (0.4,4.4) -- (0.4,6);
    \draw (0.8,4.4) -- (0.8,6);
    \draw (1.2,4.4) -- (1.2,6);
    \draw (1.6,4.4) -- (1.6,6);
    \draw (2.0,4.4) -- (2.0,6);
    \draw (2.4,4.4) -- (2.4,6);
    \draw (2.8,4.4) -- (2.8,6);
    \draw (3.2,4.4) -- (3.2,6);
    \draw (3.6,4.4) -- (3.6,6);
    \draw (4,4.4) -- (4,6);
    \node at (0.2,4.6) {1};
    \node at (0.2,5) {1};
    \node at (0.2,5.4) {1};

    \node at (0.6,4.6) {11};

    \node at (1.0,4.6) {1};
    \node at (1.0,5) {2};
    \node at (1.0,5.4) {1};

    \node at (1.4,4.6) {1};
    \node at (1.4,5) {2};
    \node at (1.4,5.4) {1};
    
    \node at (1.8,4.6) {3};
    \node at (1.8,5) {1};
    \node at (1.8,5.4) {1};

    \node at (2.2,4.6) {3};
    \node at (2.2,5) {2};
    \node at (2.2,5.4) {1};

    \node at (2.6,4.6) {1};
    \node at (2.6,5) {1};
    \node at (2.6,5.4) {1};
    \node at (2.6,5.8) {1};

    \node at (3.0,4.6) {1};
    \node at (3.0,5) {2};
    \node at (3.0,5.4) {1};
    \node at (3.0,5.8) {1};

    \node at (3.4,4.6) {1};
    \node at (3.4,5) {2};
    \node at (3.4,5.4) {1};

    \node at (3.8,4.6) {11};

    \node at (4.2,4.6) {1};
    \node at (4.2,5) {1};
    \node at (4.2,5.4) {1};
\end{tikzpicture}}}\hfil
            \subfloat[OR gadget\label{fig:OR Gadget}]{\resizebox{0.32\textwidth}{!}{\begin{tikzpicture}
    \node at (0.2,2.2) {\textcolor{black}{$\Bar{v}_1$}};
    \node at (1.4,2.2) {\textcolor{black}{$v_1$}};
    \node at (4.2,2.2) {\textcolor{black}{$\Bar{v}'$}};
    \node at (3,2.2) {\textcolor{black}{$v'$}};

    \node at (2.2,0.2) {\textcolor{black}{$\Bar{v}_2$}};
    \node at (2.2,1.4) {\textcolor{black}{$v_2$}};

    \draw[step=0.4cm] (0,0) grid (4.4,4.4);

    \draw[line width=0.5mm] (-0.9,0) -- (4.4,0);
    \draw[line width=0.5mm] (-0.9,4.4) -- (4.4,4.4);
    \draw[line width=0.5mm] (-0.9,0) -- (-0.9,4.4);
    \draw (0,0.4) -- (-0.9,0.4);
    \draw (0,0.8) -- (-0.9,0.8);
    \draw (0,1.2) -- (-0.9,1.2);
    \draw (0,1.6) -- (-0.9,1.6);
    \draw (0,2.0) -- (-0.9,2.0);
    \draw (0,2.4) -- (-0.9,2.4);
    \draw (0,2.8) -- (-0.9,2.8);
    \draw (0,3.2) -- (-0.9,3.2);
    \draw (0,3.6) -- (-0.9,3.6);
    \draw (0,4.0) -- (-0.9,4.0);
    \node at (-0.15,0.2) {1};
    \node at (-0.45,0.2) {1};
    \node at (-0.75,0.2) {1};
    \node at (-0.30,0.6) {11};
    \node at (-0.15,1.0) {1};
    \node at (-0.45,1.0) {2};
    \node at (-0.75,1.0) {1};
    \node at (-0.15,1.4) {1};
    \node at (-0.45,1.4) {2};
    \node at (-0.75,1.4) {1};
    \node at (-0.15,1.8) {1};
    \node at (-0.45,1.8) {1};
    \node at (-0.75,1.8) {1};
    \node at (-0.15,2.2) {3};
    \node at (-0.45,2.2) {1};
    \node at (-0.75,2.2) {3};
    \node at (-0.15,2.6) {3};
    \node at (-0.45,2.6) {1};
    \node at (-0.75,2.6) {3};
    \node at (-0.15,3.0) {1};
    \node at (-0.45,3.0) {4};
    \node at (-0.75,3.0) {1};
    \node at (-0.15,3.4) {1};
    \node at (-0.45,3.4) {1};
    \node at (-0.3,3.8) {11};
    \node at (-0.15,4.2) {1};
    \node at (-0.45,4.2) {1};
    
    \draw[line width=0.5mm] (0,0) -- (0,6);
    \draw[line width=0.5mm] (4.4,0) -- (4.4,6);
    \draw[line width=0.5mm] (0,6) -- (4.4,6);
    \draw (0.4,4.4) -- (0.4,6);
    \draw (0.8,4.4) -- (0.8,6);
    \draw (1.2,4.4) -- (1.2,6);
    \draw (1.6,4.4) -- (1.6,6);
    \draw (2.0,4.4) -- (2.0,6);
    \draw (2.4,4.4) -- (2.4,6);
    \draw (2.8,4.4) -- (2.8,6);
    \draw (3.2,4.4) -- (3.2,6);
    \draw (3.6,4.4) -- (3.6,6);
    \draw (4,4.4) -- (4,6);

    \node at (0.2,4.6) {1};
    \node at (0.2,5) {1};
    \node at (0.2,5.4) {1};

    \node at (0.6,4.6) {11};

    \node at (1.0,4.6) {1};
    \node at (1.0,5) {2};
    \node at (1.0,5.4) {1};
    
    \node at (1.4,4.6) {1};
    \node at (1.4,5) {1};
    \node at (1.4,5.4) {1};
    \node at (1.4,5.8) {1};

    \node at (1.8,4.6) {3};
    \node at (1.8,5) {1};
    \node at (1.8,5.4) {1};

    \node at (2.2,4.6) {3};
    \node at (2.2,5) {3};
    \node at (2.2,5.4) {1};

    \node at (2.6,4.6) {1};
    \node at (2.6,5) {1};
    \node at (2.6,5.4) {1};
    \node at (2.6,5.8) {1};

    \node at (3.0,4.6) {1};
    \node at (3.0,5) {1};
    \node at (3.0,5.4) {1};
    \node at (3.0,5.8) {1};

    \node at (3.4,4.6) {1};
    \node at (3.4,5) {2};
    \node at (3.4,5.4) {1};

    \node at (3.8,4.6) {11};

    \node at (4.2,4.6) {1};
    \node at (4.2,5) {1};
    \node at (4.2,5.4) {1};
\end{tikzpicture}}}\hfil
            \subfloat[Wire gadget\label{fig:Wire Gadget}]{\resizebox{0.32\textwidth}{!}{\begin{tikzpicture}
    \node at (0.2,2.2) {\textcolor{black}{$\Bar{v}$}};
    \node at (1.0,2.2) {\textcolor{black}{$v$}};
    \node at (4.2,2.2) {\textcolor{black}{$\Bar{v}'$}};
    \node at (3.4,2.2) {\textcolor{black}{$v'$}};

    \draw[step=0.4cm] (0,0) grid (4.4,4.4);

    \draw[line width=0.5mm] (-0.9,0) -- (4.4,0);
    \draw[line width=0.5mm] (-0.9,4.4) -- (4.4,4.4);
    \draw[line width=0.5mm] (-0.9,0) -- (-0.9,4.4);
    \draw (0,0.4) -- (-0.9,0.4);
    \draw (0,0.8) -- (-0.9,0.8);
    \draw (0,1.2) -- (-0.9,1.2);
    \draw (0,1.6) -- (-0.9,1.6);
    \draw (0,2.0) -- (-0.9,2.0);
    \draw (0,2.4) -- (-0.9,2.4);
    \draw (0,2.8) -- (-0.9,2.8);
    \draw (0,3.2) -- (-0.9,3.2);
    \draw (0,3.6) -- (-0.9,3.6);
    \draw (0,4.0) -- (-0.9,4.0);
    % Row Descriptions
    \node at (-0.15,0.2) {1};
    \node at (-0.45,0.2) {1};

    \node at (-0.30,0.6) {11};

    \node at (-0.15,1.0) {1};
    \node at (-0.45,1.0) {1};

    \node at (-0.15,1.4) {1};
    \node at (-0.45,1.4) {1};

    \node at (-0.15,1.8) {2};
    \node at (-0.45,1.8) {4};
    \node at (-0.75,1.8) {2};

    \node at (-0.15,2.2) {2};
    \node at (-0.45,2.2) {4};
    \node at (-0.75,2.2) {2};

    \node at (-0.15,2.6) {1};
    \node at (-0.45,2.6) {1};

    \node at (-0.15,3.0) {1};
    \node at (-0.45,3.0) {1};

    \node at (-0.15,3.4) {1};
    \node at (-0.45,3.4) {1};

    \node at (-0.3,3.8) {11};

    \node at (-0.15,4.2) {1};
    \node at (-0.45,4.2) {1};

    \draw[line width=0.5mm] (0,0) -- (0,5.6);
    \draw[line width=0.5mm] (4.4,0) -- (4.4,5.6);
    \draw[line width=0.5mm] (0,5.6) -- (4.4,5.6);
    \draw (0.4,4.4) -- (0.4,5.6);
    \draw (0.8,4.4) -- (0.8,5.6);
    \draw (1.2,4.4) -- (1.2,5.6);
    \draw (1.6,4.4) -- (1.6,5.6);
    \draw (2.0,4.4) -- (2.0,5.6);
    \draw (2.4,4.4) -- (2.4,5.6);
    \draw (2.8,4.4) -- (2.8,5.6);
    \draw (3.2,4.4) -- (3.2,5.6);
    \draw (3.6,4.4) -- (3.6,5.6);
    \draw (4,4.4) -- (4,5.6);
    % Column Descriptions
    \node at (0.2,4.6) {1};
    \node at (0.2,5) {1};
    \node at (0.2,5.4) {1};

    \node at (0.6,4.6) {11};

    \node at (1.0,4.6) {1};
    \node at (1.0,5) {1};
    \node at (1.0,5.4) {1};

    \node at (1.4,4.6) {1};
    \node at (1.4,5) {1};
    \node at (1.4,5.4) {1};

    \node at (1.8,4.6) {1};
    \node at (1.8,5) {2};
    \node at (1.8,5.4) {1};

    \node at (2.2,4.6) {1};
    \node at (2.2,5) {2};
    \node at (2.2,5.4) {1};

    \node at (2.6,4.6) {1};
    \node at (2.6,5) {2};
    \node at (2.6,5.4) {1};

    \node at (3.0,4.6) {1};
    \node at (3.0,5) {1};
    \node at (3.0,5.4) {1};

    \node at (3.4,4.6) {1};
    \node at (3.4,5) {1};
    \node at (3.4,5.4) {1};

    \node at (3.8,4.6) {11};

    \node at (4.2,4.6) {1};
    \node at (4.2,5) {1};
    \node at (4.2,5.4) {1};
\end{tikzpicture}}}\hfil
            \subfloat[Crossover gadget\label{fig:Crossover Gadget}]{\resizebox{0.32\textwidth}{!}{\begin{tikzpicture}
    \node at (0.2,2.2) {\textcolor{black}{$\Bar{v}_1$}};
    \node at (1.0,2.2) {\textcolor{black}{$v_1$}};
    \node at (4.2,2.2) {\textcolor{black}{$\Bar{v}_1'$}};
    \node at (3.4,2.2) {\textcolor{black}{$v_1'$}};

    \node at (2.2,0.2) {\textcolor{black}{$\Bar{v}_2'$}};
    \node at (2.2,1.0) {\textcolor{black}{$v_2'$}};
    \node at (2.2,4.2) {\textcolor{black}{$\Bar{v}_2$}};
    \node at (2.2,3.4) {\textcolor{black}{$v_2$}};

    \draw[step=0.4cm] (0,0) grid (4.4,4.4);

    \draw[line width=0.5mm] (-0.9,0) -- (4.4,0);
    \draw[line width=0.5mm] (-0.9,4.4) -- (4.4,4.4);
    \draw[line width=0.5mm] (-0.9,0) -- (-0.9,4.4);
    \draw (0,0.4) -- (-0.9,0.4);
    \draw (0,0.8) -- (-0.9,0.8);
    \draw (0,1.2) -- (-0.9,1.2);
    \draw (0,1.6) -- (-0.9,1.6);
    \draw (0,2.0) -- (-0.9,2.0);
    \draw (0,2.4) -- (-0.9,2.4);
    \draw (0,2.8) -- (-0.9,2.8);
    \draw (0,3.2) -- (-0.9,3.2);
    \draw (0,3.6) -- (-0.9,3.6);
    \draw (0,4.0) -- (-0.9,4.0);
    % Row Descriptions
    \node at (-0.15,0.2) {1};
    \node at (-0.45,0.2) {1};
    \node at (-0.75,0.2) {1};

    \node at (-0.30,0.6) {11};

    \node at (-0.15,1.0) {1};
    \node at (-0.45,1.0) {1};
    \node at (-0.75,1.0) {1};

    \node at (-0.15,1.4) {1};
    \node at (-0.45,1.4) {1};
    \node at (-0.75,1.4) {1};

    \node at (-0.15,1.8) {2};
    \node at (-0.45,1.8) {4};
    \node at (-0.75,1.8) {2};

    \node at (-0.15,2.2) {2};
    \node at (-0.45,2.2) {4};
    \node at (-0.75,2.2) {2};

    \node at (-0.15,2.6) {1};
    \node at (-0.45,2.6) {2};
    \node at (-0.75,2.6) {1};

    \node at (-0.15,3.0) {1};
    \node at (-0.45,3.0) {1};
    \node at (-0.75,3.0) {1};

    \node at (-0.15,3.4) {1};
    \node at (-0.45,3.4) {1};
    \node at (-0.75,3.4) {1};

    \node at (-0.3,3.8) {11};

    \node at (-0.15,4.2) {1};
    \node at (-0.45,4.2) {1};
    \node at (-0.75,4.2) {1};

    \draw[line width=0.5mm] (0,0) -- (0,5.6);
    \draw[line width=0.5mm] (4.4,0) -- (4.4,5.6);
    \draw[line width=0.5mm] (0,5.6) -- (4.4,5.6);
    \draw (0.4,4.4) -- (0.4,5.6);
    \draw (0.8,4.4) -- (0.8,5.6);
    \draw (1.2,4.4) -- (1.2,5.6);
    \draw (1.6,4.4) -- (1.6,5.6);
    \draw (2.0,4.4) -- (2.0,5.6);
    \draw (2.4,4.4) -- (2.4,5.6);
    \draw (2.8,4.4) -- (2.8,5.6);
    \draw (3.2,4.4) -- (3.2,5.6);
    \draw (3.6,4.4) -- (3.6,5.6);
    \draw (4,4.4) -- (4,5.6);
    % Column Descriptions
    \node at (0.2,4.6) {1};
    \node at (0.2,5) {1};
    \node at (0.2,5.4) {1};

    \node at (0.6,4.6) {11};

    \node at (1.0,4.6) {1};
    \node at (1.0,5) {1};
    \node at (1.0,5.4) {1};

    \node at (1.4,4.6) {1};
    \node at (1.4,5) {1};
    \node at (1.4,5.4) {1};

    \node at (1.8,4.6) {2};
    \node at (1.8,5) {4};
    \node at (1.8,5.4) {2};

    \node at (2.2,4.6) {2};
    \node at (2.2,5) {4};
    \node at (2.2,5.4) {2};

    \node at (2.6,4.6) {1};
    \node at (2.6,5) {2};
    \node at (2.6,5.4) {1};

    \node at (3.0,4.6) {1};
    \node at (3.0,5) {1};
    \node at (3.0,5.4) {1};

    \node at (3.4,4.6) {1};
    \node at (3.4,5) {1};
    \node at (3.4,5.4) {1};

    \node at (3.8,4.6) {11};

    \node at (4.2,4.6) {1};
    \node at (4.2,5) {1};
    \node at (4.2,5.4) {1};
\end{tikzpicture}}}\hfil
            \subfloat[Splitter gadget\label{fig:Splitter Gadget}]{\resizebox{0.32\textwidth}{!}{\begin{tikzpicture}
    \node at (0.2,2.2) {\textcolor{black}{$\Bar{v}$}};
    \node at (1.4,2.2) {\textcolor{black}{$v$}};

    \node at (4.2,2.2) {\textcolor{black}{$\Bar{v}'$}};
    \node at (3,2.2) {\textcolor{black}{$v'$}};

    \node at (2.2,0.2) {\textcolor{black}{$\Bar{v}''$}};
    \node at (2.2,1.4) {\textcolor{black}{$v''$}};

    \draw[step=0.4cm] (0,0) grid (4.4,4.4);

    \draw[line width=0.5mm] (-1.2,0) -- (4.4,0);
    \draw[line width=0.5mm] (-1.2,4.4) -- (4.4,4.4);
    \draw[line width=0.5mm] (-1.2,0) -- (-1.2,4.4);
    \draw (0,0.4) -- (-1.2,0.4);
    \draw (0,0.8) -- (-1.2,0.8);
    \draw (0,1.2) -- (-1.2,1.2);
    \draw (0,1.6) -- (-1.2,1.6);
    \draw (0,2.0) -- (-1.2,2.0);
    \draw (0,2.4) -- (-1.2,2.4);
    \draw (0,2.8) -- (-1.2,2.8);
    \draw (0,3.2) -- (-1.2,3.2);
    \draw (0,3.6) -- (-1.2,3.6);
    \draw (0,4.0) -- (-1.2,4.0);

    % Row Descriptions
    \node at (-0.15,0.2) {1};
    \node at (-0.45,0.2) {1};
    \node at (-0.75,0.2) {1};

    \node at (-0.30,0.6) {11};

    \node at (-0.15,1.0) {1};
    \node at (-0.45,1.0) {2};
    \node at (-0.75,1.0) {1};

    \node at (-0.15,1.4) {1};
    \node at (-0.45,1.4) {2};
    \node at (-0.75,1.4) {1};
    \node at (-1.05,1.4) {1};

    \node at (-0.15,1.8) {1};
    \node at (-0.45,1.8) {2};
    \node at (-0.75,1.8) {1};
    \node at (-1.05,1.8) {1};

    \node at (-0.15,2.2) {3};
    \node at (-0.45,2.2) {1};
    \node at (-0.75,2.2) {3};

    \node at (-0.15,2.6) {3};
    \node at (-0.45,2.6) {1};
    \node at (-0.75,2.6) {3};

    \node at (-0.15,3.0) {1};
    \node at (-0.45,3.0) {4};
    \node at (-0.75,3.0) {1};

    \node at (-0.15,3.4) {1};
    \node at (-0.45,3.4) {1};

    \node at (-0.3,3.8) {11};

    \node at (-0.15,4.2) {1};
    \node at (-0.45,4.2) {1};

    \draw[line width=0.5mm] (0,0) -- (0,6);
    \draw[line width=0.5mm] (4.4,0) -- (4.4,6);
    \draw[line width=0.5mm] (0,6) -- (4.4,6);
    \draw (0.4,4.4) -- (0.4,6);
    \draw (0.8,4.4) -- (0.8,6);
    \draw (1.2,4.4) -- (1.2,6);
    \draw (1.6,4.4) -- (1.6,6);
    \draw (2.0,4.4) -- (2.0,6);
    \draw (2.4,4.4) -- (2.4,6);
    \draw (2.8,4.4) -- (2.8,6);
    \draw (3.2,4.4) -- (3.2,6);
    \draw (3.6,4.4) -- (3.6,6);
    \draw (4,4.4) -- (4,6);

    % Column Descriptions
    \node at (0.2,4.6) {1};
    \node at (0.2,5) {1};
    \node at (0.2,5.4) {1};

    \node at (0.6,4.6) {11};

    \node at (1.0,4.6) {1};
    \node at (1.0,5) {2};
    \node at (1.0,5.4) {1};

    \node at (1.4,4.6) {1};
    \node at (1.4,5) {2};
    \node at (1.4,5.4) {1};
    \node at (1.4,5.8) {1};

    \node at (1.8,4.6) {3};
    \node at (1.8,5) {1};
    \node at (1.8,5.4) {1};

    \node at (2.2,4.6) {3};
    \node at (2.2,5) {3};
    \node at (2.2,5.4) {1};
    
    \node at (2.6,4.6) {1};
    \node at (2.6,5) {2};
    \node at (2.6,5.4) {1};
    \node at (2.6,5.8) {1};

    \node at (3.0,4.6) {1};
    \node at (3.0,5) {2};
    \node at (3.0,5.4) {1};
    \node at (3.0,5.8) {1};

    \node at (3.4,4.6) {1};
    \node at (3.4,5) {2};
    \node at (3.4,5.4) {1};

    \node at (3.8,4.6) {11};

    \node at (4.2,4.6) {1};
    \node at (4.2,5) {1};
    \node at (4.2,5.4) {1};
\end{tikzpicture}}}\hfil
            \subfloat[Input terminal\label{fig:Input Gadget}]{\resizebox{0.32\textwidth}{!}{\begin{tikzpicture}
    \node at (4.2,2.2) {\textcolor{black}{$\Bar{v}$}};
    \node at (3.4,2.2) {\textcolor{black}{$v$}};

    \draw[step=0.4cm] (0,0) grid (4.4,4.4);

    \draw[line width=0.5mm] (-0.9,0) -- (4.4,0);
    \draw[line width=0.5mm] (-0.9,4.4) -- (4.4,4.4);
    \draw[line width=0.5mm] (-0.9,0) -- (-0.9,4.4);
    \draw (0,0.4) -- (-0.9,0.4);
    \draw (0,0.8) -- (-0.9,0.8);
    \draw (0,1.2) -- (-0.9,1.2);
    \draw (0,1.6) -- (-0.9,1.6);
    \draw (0,2.0) -- (-0.9,2.0);
    \draw (0,2.4) -- (-0.9,2.4);
    \draw (0,2.8) -- (-0.9,2.8);
    \draw (0,3.2) -- (-0.9,3.2);
    \draw (0,3.6) -- (-0.9,3.6);
    \draw (0,4.0) -- (-0.9,4.0);

    % Row Descriptions
    \node at (-0.15,0.2) {1};
    \node at (-0.45,0.2) {1};

    \node at (-0.30,0.6) {11};

    \node at (-0.15,1.0) {1};
    \node at (-0.45,1.0) {1};

    \node at (-0.15,1.4) {1};
    \node at (-0.45,1.4) {1};

    \node at (-0.15,1.8) {2};
    \node at (-0.45,1.8) {1};
    \node at (-0.75,1.8) {1};
    
    \node at (-0.15,2.2) {2};
    \node at (-0.45,2.2) {1};
    \node at (-0.75,2.2) {1};

    \node at (-0.15,2.6) {1};
    \node at (-0.45,2.6) {1};

    \node at (-0.15,3.0) {1};
    \node at (-0.45,3.0) {1};

    \node at (-0.15,3.4) {1};
    \node at (-0.45,3.4) {1};

    \node at (-0.3,3.8) {11};

    \node at (-0.15,4.2) {1};
    \node at (-0.45,4.2) {1};

    \draw[line width=0.5mm] (0,0) -- (0,5.6);
    \draw[line width=0.5mm] (4.4,0) -- (4.4,5.6);
    \draw[line width=0.5mm] (0,5.6) -- (4.4,5.6);
    \draw (0.4,4.4) -- (0.4,5.6);
    \draw (0.8,4.4) -- (0.8,5.6);
    \draw (1.2,4.4) -- (1.2,5.6);
    \draw (1.6,4.4) -- (1.6,5.6);
    \draw (2.0,4.4) -- (2.0,5.6);
    \draw (2.4,4.4) -- (2.4,5.6);
    \draw (2.8,4.4) -- (2.8,5.6);
    \draw (3.2,4.4) -- (3.2,5.6);
    \draw (3.6,4.4) -- (3.6,5.6);
    \draw (4,4.4) -- (4,5.6);

    % Column Descriptions
    \node at (0.2,4.6) {1};
    \node at (0.2,5) {1};

    \node at (0.6,4.6) {11};

    \node at (1.0,4.6) {1};
    \node at (1.0,5) {1};

    \node at (1.4,4.6) {1};
    \node at (1.4,5) {1};

    \node at (1.8,4.6) {1};
    \node at (1.8,5) {1};

    \node at (2.2,4.6) {1};
    \node at (2.2,5) {1};

    \node at (2.6,4.6) {1};
    \node at (2.6,5) {1};
    \node at (2.6,5.4) {1};

    \node at (3.0,4.6) {1};
    \node at (3.0,5) {1};
    \node at (3.0,5.4) {1};

    \node at (3.4,4.6) {1};
    \node at (3.4,5) {1};
    \node at (3.4,5.4) {1};

    \node at (3.8,4.6) {11};

    \node at (4.2,4.6) {1};
    \node at (4.2,5) {1};
    \node at (4.2,5.4) {1};
\end{tikzpicture}}}\hfil
            \subfloat[Output terminal\label{fig:Output Gadget}]{\resizebox{0.32\textwidth}{!}{\begin{tikzpicture}
    \node at (0.2,2.2) {\textcolor{black}{$\Bar{v}$}};
    \node at (1.0,2.2) {\textcolor{black}{$v$}};

    \draw[step=0.4cm] (0,0) grid (4.4,4.4);

    \draw[line width=0.5mm] (-1.2,0) -- (4.4,0);
    \draw[line width=0.5mm] (-1.2,4.4) -- (4.4,4.4);
    \draw[line width=0.5mm] (-1.2,0) -- (-1.2,4.4);
    \draw (0,0.4) -- (-1.2,0.4);
    \draw (0,0.8) -- (-1.2,0.8);
    \draw (0,1.2) -- (-1.2,1.2);
    \draw (0,1.6) -- (-1.2,1.6);
    \draw (0,2.0) -- (-1.2,2.0);
    \draw (0,2.4) -- (-1.2,2.4);
    \draw (0,2.8) -- (-1.2,2.8);
    \draw (0,3.2) -- (-1.2,3.2);
    \draw (0,3.6) -- (-1.2,3.6);
    \draw (0,4.0) -- (-1.2,4.0);

    % Row Descriptions
    \node at (-0.15,0.2) {1};
    \node at (-0.45,0.2) {1};

    \node at (-0.30,0.6) {11};

    \node at (-0.15,1.0) {1};
    \node at (-0.45,1.0) {1};

    \node at (-0.15,1.4) {1};
    \node at (-0.45,1.4) {1};

    \node at (-0.15,1.8) {1};
    \node at (-0.45,1.8) {1};
    \node at (-0.75,1.8) {2};

    \node at (-0.15,2.2) {1};
    \node at (-0.45,2.2) {1};
    \node at (-0.75,2.2) {2};

    \node at (-0.15,2.6) {1};
    \node at (-0.45,2.6) {1};

    \node at (-0.15,3.0) {1};
    \node at (-0.45,3.0) {1};

    \node at (-0.15,3.4) {1};
    \node at (-0.45,3.4) {1};

    \node at (-0.3,3.8) {11};

    \node at (-0.15,4.2) {1};
    \node at (-0.45,4.2) {1};

    \draw[line width=0.5mm] (0,0) -- (0,5.6);
    \draw[line width=0.5mm] (4.4,0) -- (4.4,5.6);
    \draw[line width=0.5mm] (0,5.6) -- (4.4,5.6);
    \draw (0.4,4.4) -- (0.4,5.6);
    \draw (0.8,4.4) -- (0.8,5.6);
    \draw (1.2,4.4) -- (1.2,5.6);
    \draw (1.6,4.4) -- (1.6,5.6);
    \draw (2.0,4.4) -- (2.0,5.6);
    \draw (2.4,4.4) -- (2.4,5.6);
    \draw (2.8,4.4) -- (2.8,5.6);
    \draw (3.2,4.4) -- (3.2,5.6);
    \draw (3.6,4.4) -- (3.6,5.6);
    \draw (4,4.4) -- (4,5.6);
    
    % Column Descriptions
    \node at (0.2,4.6) {1};
    \node at (0.2,5) {1};
    \node at (0.2,5.4) {1};

    \node at (0.6,4.6) {11};

    \node at (1.0,4.6) {1};
    \node at (1.0,5) {1};
    \node at (1.0,5.4) {1};

    \node at (1.4,4.6) {1};
    \node at (1.4,5) {1};
    \node at (1.4,5.4) {1};

    \node at (1.8,4.6) {1};
    \node at (1.8,5) {1};
    \node at (1.8,5.4) {1};

    \node at (2.2,4.6) {1};
    \node at (2.2,5) {1};

    \node at (2.6,4.6) {1};
    \node at (2.6,5) {1};
        \node at (3.0,4.6) {1};
    \node at (3.0,5) {1};

    \node at (3.4,4.6) {1};
    \node at (3.4,5) {1};

    \node at (3.8,4.6) {11};

    \node at (4.2,4.6) {1};
    \node at (4.2,5) {1};
\end{tikzpicture}}}
        \caption{Nonogram gadgets}
        \label{fig:Gadgets}
    \end{figure}
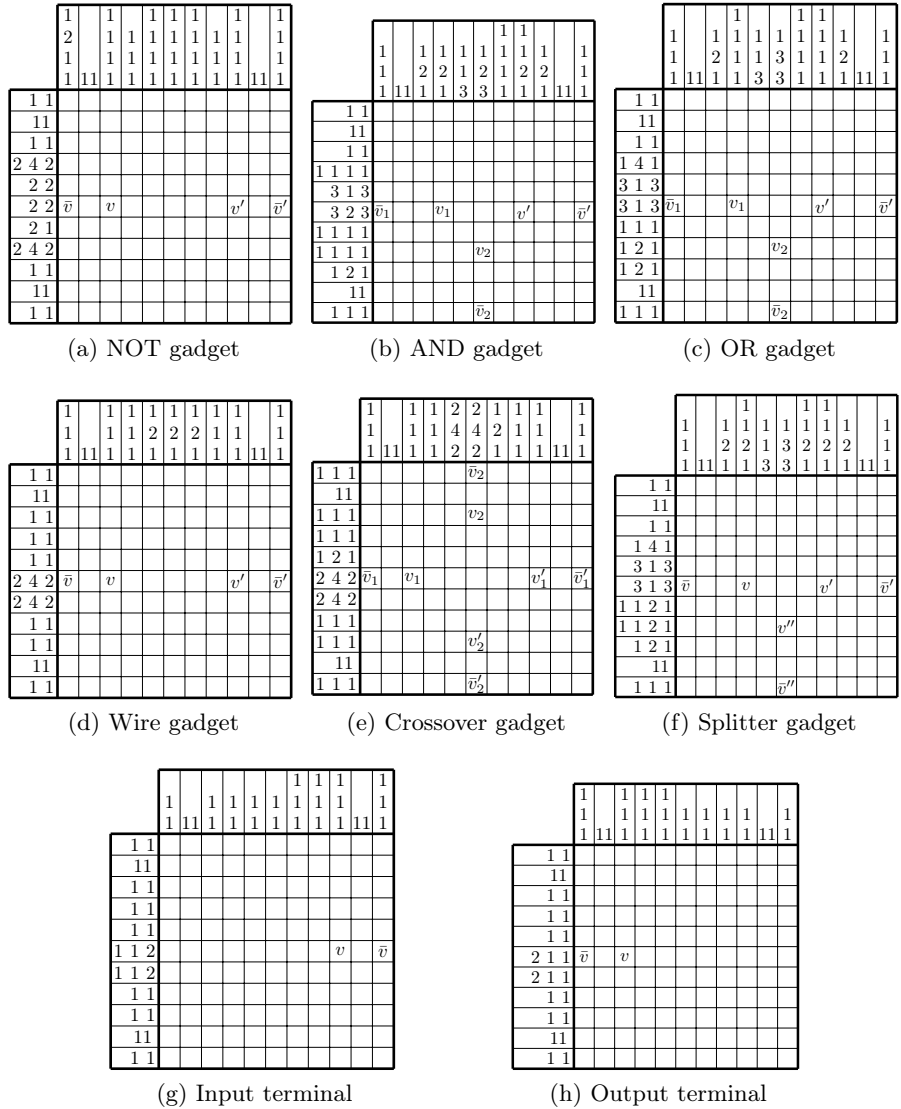

    Note that the size of the construction is polynomial in the size of $\phi$ and can be constructed in time polynomial in the size of $C_{\psi}$. Next, it must be shown that $\phi$ is 
    unsatisfiable if and only if an inference is possible for the Nonogram construction. It suffices to show that $C_{\psi}$ is unsatisfiable if and only if an inference is possible for the
    Nonogram construction.
    \begin{enumerate}
        \item Suppose $C_{\psi}$ is unsatisfiable. Then, for every assignment of values to the variable inputs of $C_{\psi}$, the output is false. For the Nonogram board, no matter the
        fillings of the input terminals, the filled cell in the output terminal will be $v$. Thus, the Nonogram construction is in INFERENCE. 
        
        \item Suppose $C_{\psi}$ is satisfiable. Then, there must exist an assignment of values to the variable inputs of $C_{\psi}$ such that the output is true. For the Nonogram 
        construction, there is then a filling of the input terminals such that the filled cell in the output terminal will be $\Bar{v}$. However, we know that there \emph{must} exist an 
        assignment to the input variables, namely those taking $x^*$ to be false, for which the output will be false. Correspondingly, there must exist a filling of the input terminals of 
        the Nonogram construction such that the output terminal will have $v$ filled. Thus, when $C_{\psi}$ is satisfiable, the Nonogram construction is not in INFERENCE. 
    \end{enumerate}
    
\end{proof}

This completes the proof of the theorem. \qed

\section{Phase Transition}
\label{sec:Phase Transition}
Assuming P $\neq$ NP, the previous complexity result would suggest that Nonogram is not a puzzle that can be played efficiently. Nonetheless, it is a popular combinatorial puzzle enjoyed
by millions. We explain this by showing, empirically, phase transition behavior for the INFERENCE problem. Boards with filled cell density below the transition threshold almost surely
do not have an inferable cell, while the probability of an inferable cell occurring for boards with filled cell density above the threshold is almost one. We use a reduction of a Nonogram
board to Boolean unsatisfiability to ``play'' Nonogram. 

The reduction leverages the natural expressability of a Nonogram board as a regular expression. For an arbitrary description $[l_1,l_2,\dots,l_t]$ (assuming all $l_i > 0$), the regular
expression $0^*1^{l_1}0^+1^{l_2}0^+\dotsm1^{l_t}0^*$ recognizes all strings satisfying the description. It follows that a deterministic finite automaton can be constructed recognizing the
same language. Then, a Boolean formula is constructed describing the behavior of the automaton, such that the formula is satisfied if an only if the input string to the automoton is
accepted and of the proper length (given by the size of the board). The encoding for a board is a the conjunction of the Boolean formulas for the rows and columns. For a line of
length $n$ with description $0^*1^{l_1}0^+1^{l_2}0^+\dotsm1^{l_t}0^*$, there are $(5n+2)(t+1+\sum_{i=1}^t l_i) - 4$ clauses, $(14n+2)t + 8n - 2 + (11n+2)\sum_{i=1}^t l_i$ total variables,
and $(2n+1)(t+\sum_{i=1}^t l_i) + n$ distinct variables \cite{MYthesis}. Note that $t$, the number of description elements, and $\sum_{i=1}^t l_i$, the number of filled tiles, are both 
bounded by $n$. Thus, there are $O(n^2)$ variables and $O(n^2)$ clauses in the formula for an input description of a line with $n$ cells. The precise encoding algorithms developed can be 
found in \cite{MYthesis}.

\begin{figure}
    \centering
    \subfloat[Empty board\label{fig:Empty}]{
        \begin{tikzpicture}
            \draw[step=0.5cm] (0,0) grid (2.5,2.5);
        \end{tikzpicture}
    }\hfil
    \subfloat[Filled randomly\label{fig:Filled}]{
        \begin{tikzpicture}
            \draw[step=0.5cm] (0,0) grid (2.5,2.5);
            
            \filldraw[draw=black,color=black] (0,0) rectangle (0.5,0.5);
            \filldraw[draw=black,color=black] (1.0,0) rectangle (1.5,0.5);
            
            \filldraw[draw=black,color=black] (0.5,0.5) rectangle (1.0,1.0);
            \filldraw[draw=black,color=black] (1.0,0.5) rectangle (1.5,1.0);

            \filldraw[draw=black,color=black] (0.5,1.5) rectangle (1.0,2.0);
            \filldraw[draw=black,color=black] (1.5,1.5) rectangle (2.0,2.0);

            \filldraw[draw=black,color=black] (1.0,2.0) rectangle (1.5,2.5);
        \end{tikzpicture}
    }\hfil
    \subfloat[Descriptions generated\label{fig:Descriptions}]{
        \begin{tikzpicture}
            \draw[step=0.5cm] (0,0) grid (2.5,2.5);
            
            \draw[line width=0.5mm] (-1.0,0) -- (2.5,0);
            \draw[line width=0.5mm] (-1.0,2.5) -- (2.5,2.5);
            \draw[line width=0.5mm] (-1.0,0) -- (-1.0,2.5);

            \draw[line width=0.5mm] (0,0) -- (0,3.5);
            \draw[line width=0.5mm] (2.5,0) -- (2.5,3.5);
            \draw[line width=0.5mm] (0,3.5) -- (2.5,3.5);
            
            \draw (0,0.5) -- (-1.0,0.5);
            \draw (0,1.0) -- (-1.0,1.0);
            \draw (0,1.5) -- (-1.0,1.5);
            \draw (0,2.0) -- (-1.0,2.0);

            \draw (0.5,2.5) -- (0.5,3.5);
            \draw (1.0,2.5) -- (1.0,3.5);
            \draw (1.5,2.5) -- (1.5,3.5);
            \draw (2.0,2.5) -- (2.0,3.5);

            % Row Descriptions
            \node at (-0.25,0.25) {1};
            \node at (-0.75,0.25) {1};
    
            \node at (-0.25,0.75) {2};

            \node at (-0.25,1.75) {1};
            \node at (-0.75,1.75) {1};

            \node at (-0.25,2.25) {1};

            % Column Descriptions
            \node at (0.25,2.75) {1};
    
            \node at (0.75,2.75) {1};
            \node at (0.75,3.25) {1};

            \node at (1.25,2.75) {2};
            \node at (1.25,3.25) {1};

            \node at (1.75,2.75) {1};
        \end{tikzpicture}
    }
    
    \caption{Board generation process}
    \label{fig:Board Generation}
\end{figure}
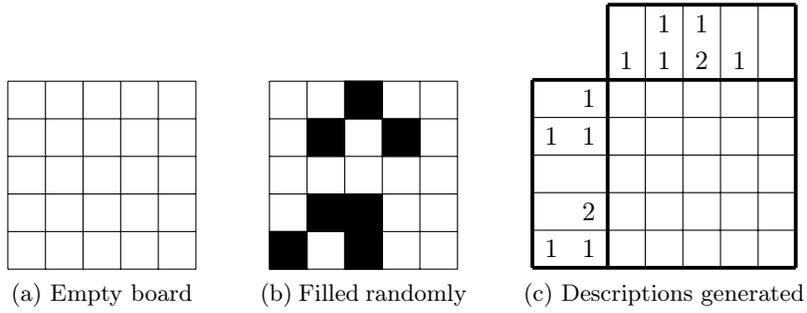

Before solving, boards must be generated. As noted in literature, one must take care when defining a ``random'' instance of a problem \cite{FrancoPaul83}. For this analysis, only boards 
that are consistent are considered. To generate a random consistent board with filled cell density $\rho$ and dimension $N \times N$, a random number from zero to one is sampled uniformly 
for each cell. If the number is less than or equal to $\rho$, the cell is filled. Otherwise it is left as empty. Boards generated with filled cell density $\rho$ will have $\rho N^2$ 
filled cells in expectation. The descriptions associated with this filling can then be extracted to create a puzzle. As noted previously, this ensures that every board created has a 
solution. With a partially filled board generated, descriptions for each row and column can be created by finding the runs that occur in each of the rows and columns, in order. By 
convention, a column starts at the top and a row starts on the left (Fig. \ref{fig:Descriptions}). With the descriptions, the traditional Nonogram board has been created 
and is ready for encoding as a Boolean formula.

The algorithm to test the boards for inferability is as follows:
\begin{enumerate}
    \item Generate filled board $B$ as a 0-1 matrix
    \item Extract descriptions for each row and column
    \item Encode each row and column as a Boolean formula, conjuncting to obtain $\phi$ satisfied if and only if every row and column of $B$ is satisfied
    \item Set accumulator \emph{inferredFilled} = 0
    \item For the Boolean variable $x_{ij}$ associated with each cell $c_{ij}$:
    \begin{enumerate}
        \item Test $\psi = \Bar{x}_{ij} \land \phi$ for satisfiability
        \item If $\psi$ satisfiable do nothing
        \item Otherwise increment \emph{inferredFilled} by one
    \end{enumerate}
\end{enumerate}

\begin{figure}
    \centering
    \includegraphics[width = \textwidth]{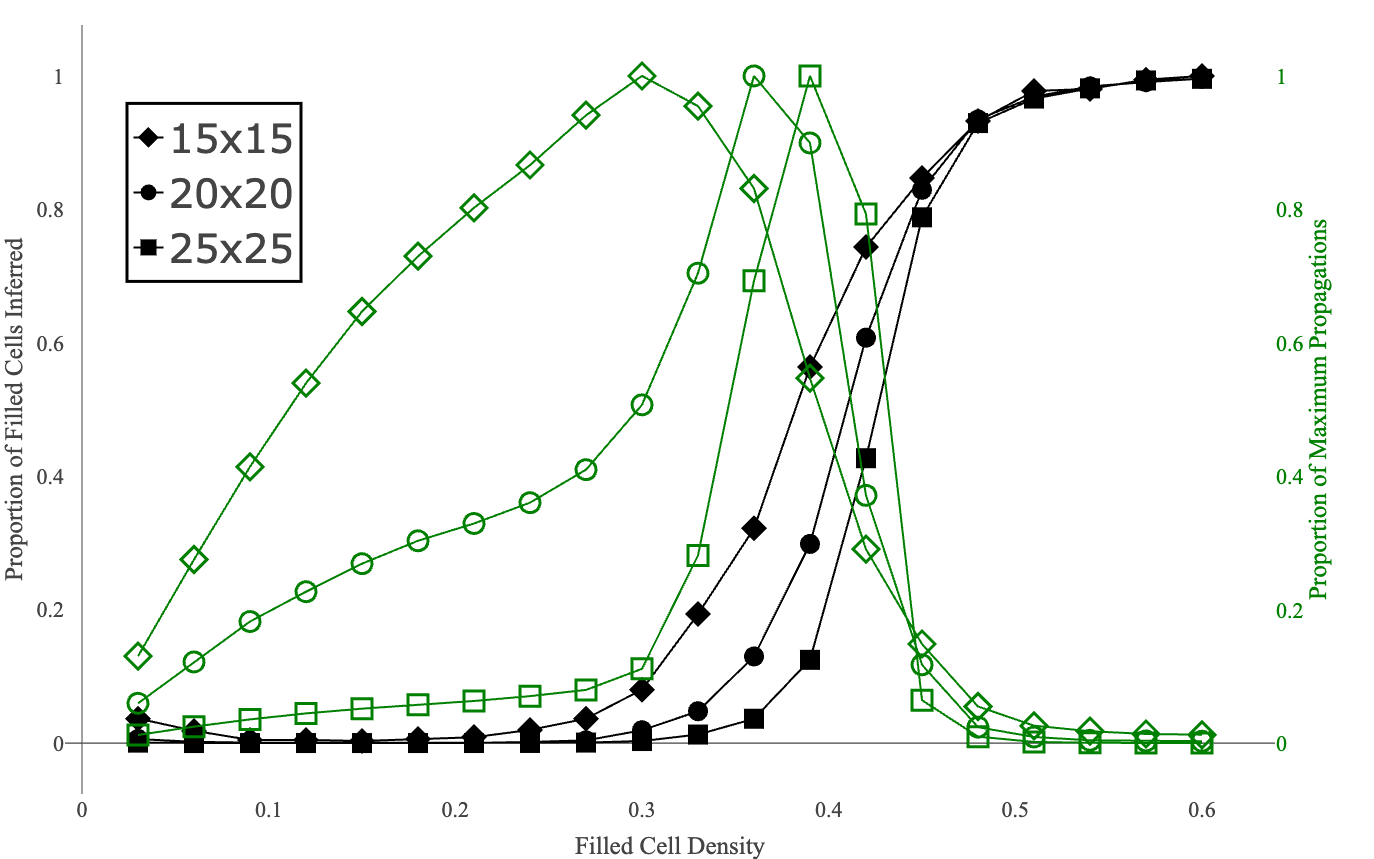}
    \caption{Phase transition behavior for Nonogram INFERENCE}
    \label{fig:The Plot}
\end{figure}

This process is repeated for each board, at each density, for each board size. For each density, 250 boards are generated. Due to computational constraints, only puzzle sizes of 15x15,
20x20, and 25x25 were tested (Figure \ref{fig:The Plot}).

On the horizontal axis is the filled cell density. For each density, the average proportion of filled cells inferred is plotted in black (the left vertical axis), with each point shape 
corresponding to a different size of puzzle. As expected, at lower densities few of the filled cells can be inferred by logical inference. There appears to be a threshold between 
densities of 0.39 and 0.42 at which the average proportion of filled cells inferred quickly increases to almost one. This behavior becomes sharper as the size of puzzle increases.

It can be seen that for the 15x15 puzzles, the average proportion of filled cells actually decreases slightly from a non-zero amount at the beginning. This is the byproduct of the relatively small board size. With a filled cell density of 0.03, there are only 6.75 filled cells in expectation. As a result, on a particular board, it is possible to generate puzzles with even fewer cells, possibly creating entirely inferable puzzles. For larger dimension, the probability of generating such boards quickly drops, a pattern that can already be seen from the increase in dimension from 15x15 to 20x20 (which has 12 expected filled cells at a density of 0.03).

Also pictured is the computational effort required to solve the puzzles. Given that the underlying process is SAT solving, metrics relating to the SAT solving process are tracked. In this 
case, the average number of propagations by the solver for each board are measured. However, propagation count is strongly associated with formula size, and the recurrences for formula size are parameterized by the size of the board. As such, the difficulty measure is plotted with empty point shapes and adjusted for board size, showing the proportion of average propagations scaled by the maximum result found for that board size. With this adjustment, the relationship between filled cell density and solving difficulty can be compared for different board sizes. Average propagation count appears to peak peaks around the phase transition threshold. This makes sense, as this is the region where the status as inferable or not inferable is most uncertain. Just as with the average proportion inferred, the spike in difficulty is starker for larger boards.

Additionally, the difficulty is not symmetric across the threshold. The average propagations is much higher for densities below the threshold than it is for densities above. This is due to 
the under/over-constrained nature of the Nonogram boards below/above the threshold. For boards below the threshold, the underconstrained boards have far more candidate solutions, and thus 
many less literals are fixed. This leads to a considerable increase in average propagations. For boards above the threshold, there are far fewer solutions, and as a result fewer 
propagations need to be executed (on average) in order to determine inferability. 

\begin{figure}
    \centering
    \subfloat[Average clauses at each density\label{fig:Clauses}]{\includegraphics[width = 0.48\textwidth]{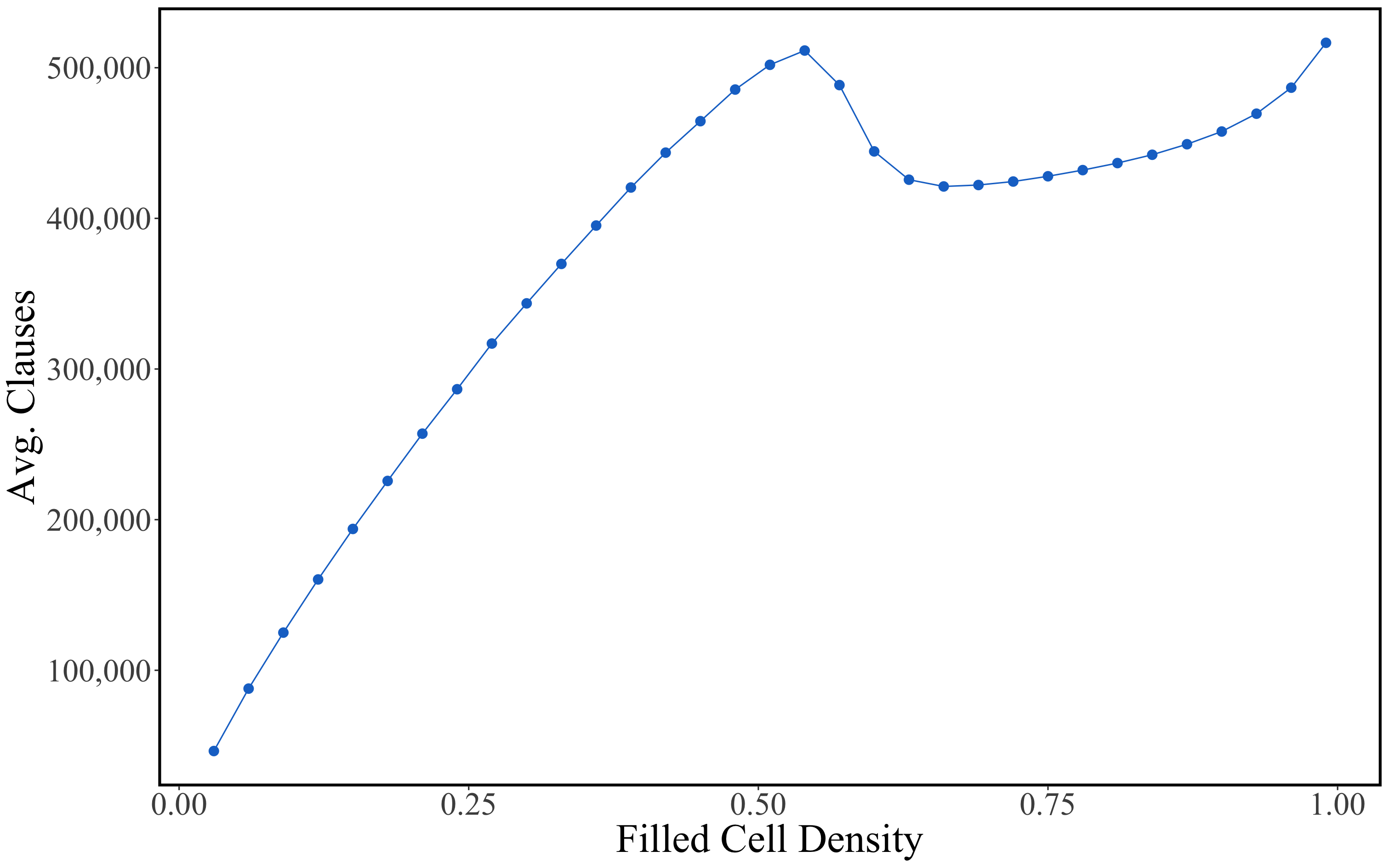}}
    \hfil
    \subfloat[Average variables at each density\label{fig:Variables}]{\includegraphics[width = 0.48\textwidth]{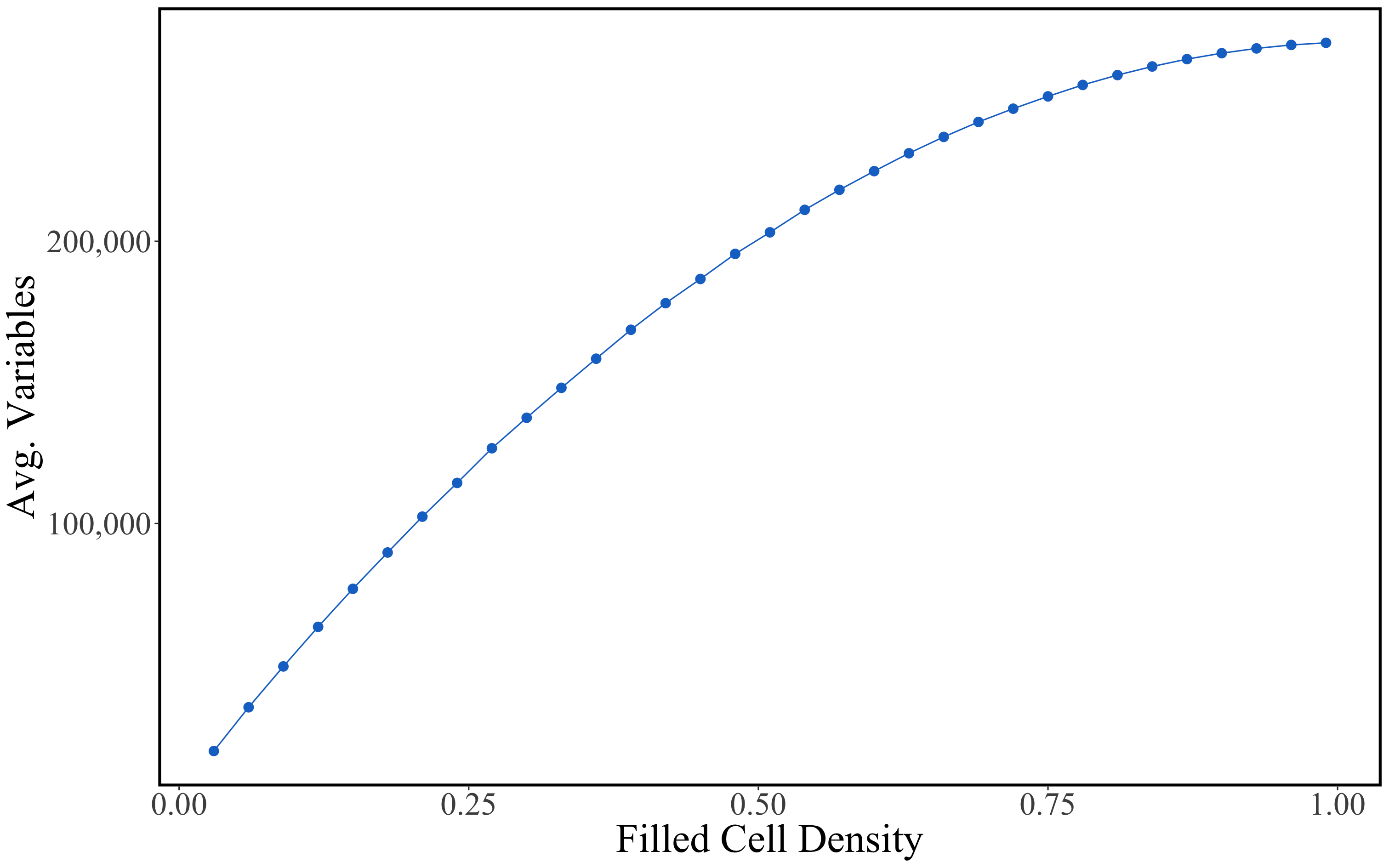}}
    \caption{Formula size across densities}
    \label{fig:Formula Size Plots}
\end{figure}

Finally, it should be noted that formula size is also driven by the number of filled cells on the board and the number of runs. One could then argue that the asymmetry of average 
propagations or spike in average propagations simply comes from the fact that the average formulas follow the pattern presented above and in fact the difficulty is uniform across 
densities. This is not the case. Per the recurrences for formula size, the formulas are larger both in terms of variables and clauses for higher filled cell densities. However, there is a 
trade-off at medium filled cell densities where the number of clauses shrinks. Since the number of clauses depends both on the number of filled cells and the number of description 
elements, there is a small dip in clauses when the descriptions start to meld together, decreasing the number of description elements considerably without much change in the number of 
filled cells in that description. This logic is borne out in generated formulas as well. Below are the average clause and variable counts across filled cell densities from zero to one, 
with each point being the average of 500 40x40 boards (Fig. \ref{fig:Formula Size Plots}).

Average clause and variable counts and computational cost appear to follow different patterns. The number of clauses and variables is greatest above a filled cell density of 0.5, the puzzles for which solving is the easiest (fewest propagations). 

\section{Conclusion}
This work builds on complexity results for the combinatorial puzzle Nonogram, considering problems directly related to playing the puzzle in a fashion that humans do. We have shown that inferring Nonogram unfilled cell values is co-NP-complete. To remedy this result with the popularity of the puzzle, we have demonstrated empirically phase transition behavior for the inference problem. In the process we have demonstrated that the hardness of the puzzle peaks at the transition, and outside of the transition threshold, the computation required to solve puzzles is manageable.

\begin{credits}
\subsubsection{\ackname} The authors would like to thank Robert Rose and Dan Licata for valuable feedback during the revision process.

\subsubsection{\discintname}
The authors have no competing interests to declare that are relevant to the content of this article.
\end{credits}
\bibliographystyle{splncs04}
\bibliography{bibliography}

\end{document}